\documentclass[aps,prl,twocolumn,superscriptaddress]{revtex4-1} 

\usepackage[utf8]{inputenc}  
\usepackage[T1]{fontenc}
\usepackage{amsmath, amssymb, amsxtra}
\usepackage{mathtools, nccmath}
\usepackage{siunitx}
\DeclareSIUnit\Erec{\ensuremath{\textit{E}_\text{rec}}}
\usepackage{graphicx,float,color,ctable}
\usepackage{csquotes}
\usepackage[english]{babel}
\addto\captionsenglish{}
\addto\captionsenglish{}

\begin{document}

\title{	A two-particle, four-mode interferometer for atoms }

\author{Pierre Dussarrat}
\author{Maxime Perrier}
\affiliation{Laboratoire Charles Fabry, Institut d'Optique Graduate School, CNRS, Université Paris‐Saclay, 91120 Palaiseau, France}

\author{Almazbek Imanaliev}
\affiliation{LNE-SYRTE, Observatoire de Paris, PSL Research University, CNRS, Sorbonne Universités, UPMC Université Paris 06, 75014 Paris, France}

\author{Raphael Lopes}
\affiliation{Cavendish Laboratory, University of Cambridge, Cambridge CB3 0HE, United Kingdom}

\author{Alain Aspect}
\author{Marc Cheneau}
\author{Denis Boiron}
\author{Christoph I. Westbrook}
\affiliation{Laboratoire Charles Fabry, Institut d'Optique Graduate School, CNRS, Université Paris‐Saclay, 91120 Palaiseau, France}

\begin{abstract}
	We present a free-space interferometer to observe two-particle interference of a pair of atoms with entangled momenta. The source of atom pairs is a Bose--Einstein condensate subject to a dynamical instability, and the interferometer is realized using Bragg diffraction on optical lattices, in the spirit of our recent Hong--Ou--Mandel experiment. We report on an observation ruling out the possibility of a purely mixed state at the input of the interferometer. We explain how our current setup can be extended to enable a test of a Bell inequality on momentum observables.
\end{abstract}

\maketitle

A key element of the second quantum revolution \cite{dowling2003quantum,Aspect:2004introductionsuqm} is entanglement \cite{Feynman:1982yx}. Its extraordinary character comes from the fact that the many-body wave-function of entangled particles can only be described in a configuration space associated with the tensor product of the configuration spaces of the individual particles. When one insists on describing it  in our ordinary space-time, one has to face the problem of non-locality \cite{gisin1995relevant, Aspect:2002quantumunsp, tichy2011essential}. This is clearly illustrated by the violation of Bell's inequalities \cite{Bell:1964cr}, which apply to any system that can be described in the spirit of the local realist worldview of Einstein, in which  physical reality lies in our ordinary space-time \cite{Einstein1935}.

While the violation of Bell's inequalities stems from two-particle interferences observed with entangled pairs, the converse is not true: not all phenomena associated with two-particle interference can lead to a violation of Bell's inequalities. This is for instance the case of the Hanbury Brown--Twiss effect for thermal bosons \cite{Fano1961, Glauber1965}, or the Hong--Ou--Mandel effect \cite{Hong1987}: the quantum description appeals to two-particle interference but no non-locality is involved. This is because the latter effects involve only two modes for two indistinguishable particles \footnote{By \enquote{mode}, we mean a single-particle wavefunction, whether in real space or some other space, which can be occupied by some number of identical particles.}, while a configuration leading to the violation of Bell's inequalities requires four modes that can be made to interfere two by two in different places
\footnote{The requirement for four modes holds for systems of two particles. In the context of continuous variables, configurations involving only two  modes can also lead to violations of Bell's inequalities (see, for instance, \cite{Wenger2003, Cavalcanti2011}).}. 

Ever more ideal experimental tests of Bell's inequalities have been performed with low energy photons, internal states of trapped ions and nitrogen-vacancy centers (see references in \cite{Aspect:1999gm, aspect2015viewpoint}). But we know of no experiments on two-particle interference in four modes associated with the motional degrees of freedom (position or momentum) of massive particles, and in a configuration permitting a Bell inequality test \footnote{A two-electron interference in four momentum modes was reported in Ref.~\cite{Waitz2016}, but without access to all four joint detection probabilities.}. Such tests involving mechanical observables are desirable, in particular because they may allow one to touch upon the interface between quantum mechanics and gravitation \cite{Penrose1996}. 

\begin{figure}
	\includegraphics{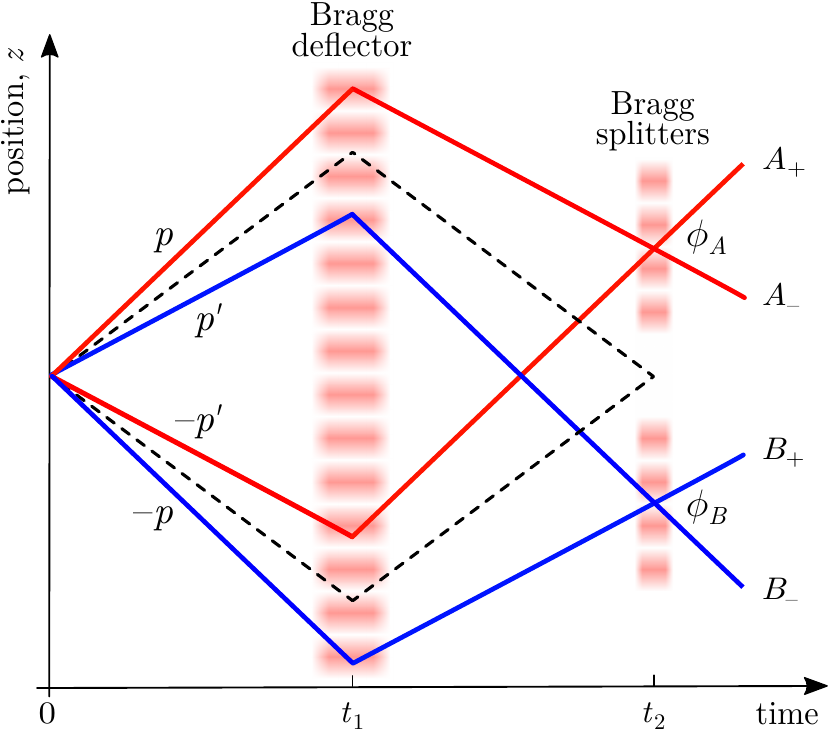}
	\caption{ \label{Fig:1}
		Diagram of a two-particle, four-mode interferometer. An atom pair in the entangled momentum state (\ref{Eq:Psi}) is emitted at time $t = 0$.
		Using Bragg diffraction on optical lattices, the four input modes are then deflected at time $t_1$, and mixed two by two at time $t_2 = 2\,t_1$ on two independent splitters $A$ and $B$, with phases $\phi_A$ and $\phi_B$.
		The interference is read out by detecting the atoms in the output modes $A_\pm$, $B_\pm$, and measuring the probabilities of joint detection $\mathcal P(A_\pm,B_\pm)$.
		The Bragg deflector and splitters differ from their optical analogs,
		because rather than reversing the incident momentum, they
		translate the momentum by a reciprocal lattice vector $\pm \hbar k_\ell$.
		The dashed lines show the Hong–Ou–Mandel configuration.
	}
\end{figure}

In this paper, we present a two-particle interferometer for momentum entangled atoms and report on an initial implementation.
To understand the experiment, consider an entangled state consisting of a pair of atoms in a superposition of distinct momentum modes labeled by $\pm p$ and $\pm p'$:
\begin{equation}
	|\Psi\rangle = \mfrac{1}{\sqrt 2} \left( |p,-p\rangle + |p',-p'\rangle \right) \; .
	\label{Eq:Psi}
\end{equation}
This superposition can be probed with the interferometer shown in Fig.~\ref{Fig:1}. An analogous interferometer for photons was proposed in Ref.~\cite{Horne:1989gi}, implemented in Ref.~\cite{Rarity1990}, and resulted in a Bell inequality violation. Similar configurations for atoms were also analyzed in Refs.~\cite{Kofler2012, Lewis-Swan2015}.
Although our results do not yet prove that we have an entangled state, they do exclude the possibility of a statistical mixture.

The input modes $p$ and $-p'$ of our interferometer are deflected and mixed on the {50:50} splitter $A$. Similarly, the input modes $p'$ and $-p$ are deflected and mixed on the {50:50} splitter $B$. The deflection and mixing are realized with Bragg diffraction on optical lattices. The deflecting lattice is common to the four input modes and is applied at time $t_1$. The splitting lattices $A$ and $B$ are applied at time $t_2 = 2\,t_1$ (the time origin is set at the instant of pair emission).
The four output modes of the interferometer, $A_\pm$ and $B_\pm$, can be written in terms of the four input modes \cite{Supplemental}:
\begin{align}
	|A_+\rangle & = \mfrac{-1}{\sqrt 2} \left( e^{-i(\phi_A-\phi_D)} |p\rangle + i\,e^{-i\phi_D} |-p'\rangle \right) \; , \label{Eq:BSA1} \\
	|A_-\rangle & = \mfrac{-1}{\sqrt 2} \left( i\,e^{i\phi_D} |p\rangle + e^{i(\phi_A-\phi_D)} |-p'\rangle \right) \; , \label{Eq:BSA2} \\
	|B_+\rangle & = \mfrac{-1}{\sqrt 2} \left( e^{-i(\phi_B-\phi_D)} |p'\rangle + i\,e^{-i\phi_D} |-p\rangle \right) \; , \label{Eq:BSB1} \\
	|B_-\rangle & = \mfrac{-1}{\sqrt 2} \left( i\,e^{i\phi_D} |p'\rangle + e^{i(\phi_B-\phi_D)} |-p\rangle \right) \; . \label{Eq:BSB2}
\end{align}
Here, the phases $\phi_D$, $\phi_A$ and $\phi_B$ are the phase differences between the laser beams forming the deflecting lattice ($\phi_D$) and the splitting lattices ($\phi_A$ and $\phi_B$); they can in principle be separately controlled. In the above equations we have omitted overall phase factors due to propagation.

Inverting equations (\ref{Eq:BSA1}–\ref{Eq:BSB2}), one readily obtains the expression of the entangled state (\ref{Eq:Psi}) at the output of the interferometer, which solely depends on $\phi_A$ and $\phi_B$:
\begin{equation}
	\label{Eq:Psi_out}
	\begin{split}
	|\Psi_\text{out}\rangle =
		\smash{\mfrac{1}{2\sqrt 2}} \Big[ \; 
			- & i \big( e^{i\phi_A} + e^{i\phi_B} \big) |A_+,B_+\rangle \\
			+ & \big( e^{i(\phi_A-\phi_B)} - 1 \big) |A_+,B_-\rangle \\
			+ & \big( e^{-i(\phi_A-\phi_B)} - 1 \big) |A_-,B_+\rangle \\
			- & i \big( e^{-i\phi_A} + e^{-i\phi_B} \big) |A_-,B_-\rangle
		\; \Big] \; .
	\end{split}
\end{equation}
The probabilities of joint detection in the output modes are given by the squared modulus of the complex amplitudes of the corresponding pair states:
\begin{align}
	\mathcal P (A_+,B_+) & = \mathcal P (A_-,B_-) = \mfrac{1}{2} \cos^2 \big[ (\phi_A-\phi_B)/2 \big] \label{Eq:PjointSym} \; , \\
	\mathcal P (A_+,B_-) & = \mathcal P (A_-,B_+) = \mfrac{1}{2} \sin^2 \big[ (\phi_A-\phi_B)/2 \big] \label{Eq:PjointAsym} \; ,
\end{align}
while the probabilities of single detection are all equal to $1/2$.
The entangled nature of the initial state is manifest in the oscillation of the joint detection probabilities as a function of the phase difference $(\phi_A-\phi_B)$.
If rather, we had initially a statistical mixture of the pair states $\vert p, -p\rangle$ and $\vert p', -p'\rangle$, there would be no modulation and the probabilities of joint detection would all be equal to $1/4$.
The four joint detection probabilities can also be combined in a single correlation coefficient:
\begin{align}
	E & \begin{multlined}[t]
			= \mathcal P (A_+,B_+) + \mathcal P (A_-,B_-) \\[1ex]
				- \mathcal P (A_+,B_-) - \mathcal P (A_-,B_+)
		\end{multlined} \\
	  & = V \cos (\phi_A-\phi_B) \; .
	\label{Eq:correl}
\end{align}
The visibility $V$ is equal to unity for the input state (\ref{Eq:Psi}), but it may be reduced in a real experiment due for example to decoherence, or the presence of additional pairs. In the case of a statistical mixture, the correlation coefficient would be equal to zero. Of course, a Bell inequality test remains possible provided $V > 1/\sqrt 2$ \cite{Clauser1969}. 

We now come to our experimental realization. A gaseous Bose–Einstein condensate (BEC) containing \num{7e4} Helium-4 atoms in the metastable $\text{2}\prescript{3}{}{\text S}_1, m_J=1$ electronic state is confined in an ellipsoidal optical trap with its long axis along the vertical ($z$) direction. The emission of atom pairs occurs in the presence of a vertical, moving optical lattice formed by the interference of two laser beams with slightly different frequencies \cite{Supplemental}.
It results from the scattering of two atoms from the BEC and can be thought of as a spontaneous, degenerate four-wave mixing process \cite{Bonneau2013}. The lattice is switched on and off adiabatically in \SI{100}{\micro\s}, and is maintained at a constant depth for \SI{600}{\micro\s}. 
The lattice hold time is tuned to produce a peak atom pair density in velocity space of about \num{3e-3} detected pairs per \si{(\mm\per\s)^3}.
The optical trap is switched off abruptly as soon as the lattice depth is returned to zero. The atoms are then transferred to the magnetically insensitive $m_J=0$ state with a two-photon Raman transition and fall freely under the sole influence of gravity. They end their fall on a micro-channel plate detector located \SI{46}{\cm} below the position of the optical trap \cite{Schellekens2005}. The detector records the impact of each atom with an efficiency $\sim \SI{25}{\percent}$. We store the arrival times and horizontal positions ($x$-$y$-plane), and reconstruct the initial three-dimensional velocity of every detected atom.

\begin{figure}
	\includegraphics{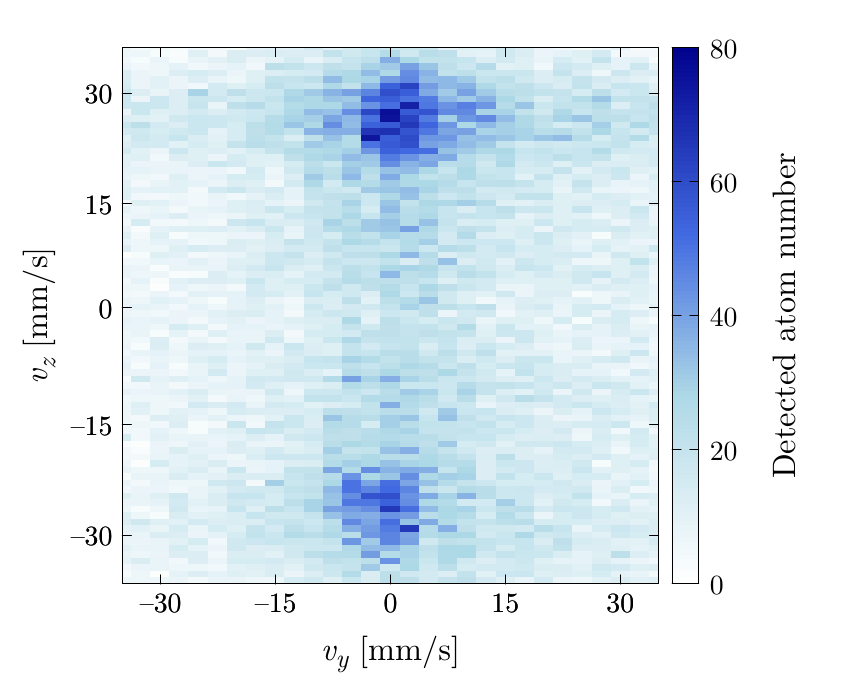}
	\caption{ \label{Fig:2}
		Initial velocity distribution of the emitted atom pairs in the $y$-$z$-plane. The color scale represents the total number of atoms detected over \num{1169} repetitions of the experiment inside 
		an integration volume of $9.2\times 2.4 \times 0.9\, (\si{\mm\per\s})^{3}$ \cite{Supplemental}.
		The velocities are defined with respect to the center-of-mass velocity of the atom pairs, which was measured to be \SIlist{0;0;94}{\mm\per\s} along the $x$, $y$ and $z$ directions, respectively.
	}
\end{figure}

In Fig.~\ref{Fig:2}, we show the initial velocity distribution of the emitted atom pairs in the $y$-$z$-plane. Here, and in the rest of the article, velocities are expressed in the center-of-mass reference frame of the free-falling pairs.  
The distribution is bimodal, and symmetric under rotation about the $z$-axis, reflecting the one-dimensional character of the pair emission. We do observe, however, a slight asymmetry in the height of the two maxima. We attribute this asymmetry to momentum-dependent losses occurring during the short time when the emitted atoms spatially overlap with the BEC.

\begin{figure}
	\includegraphics{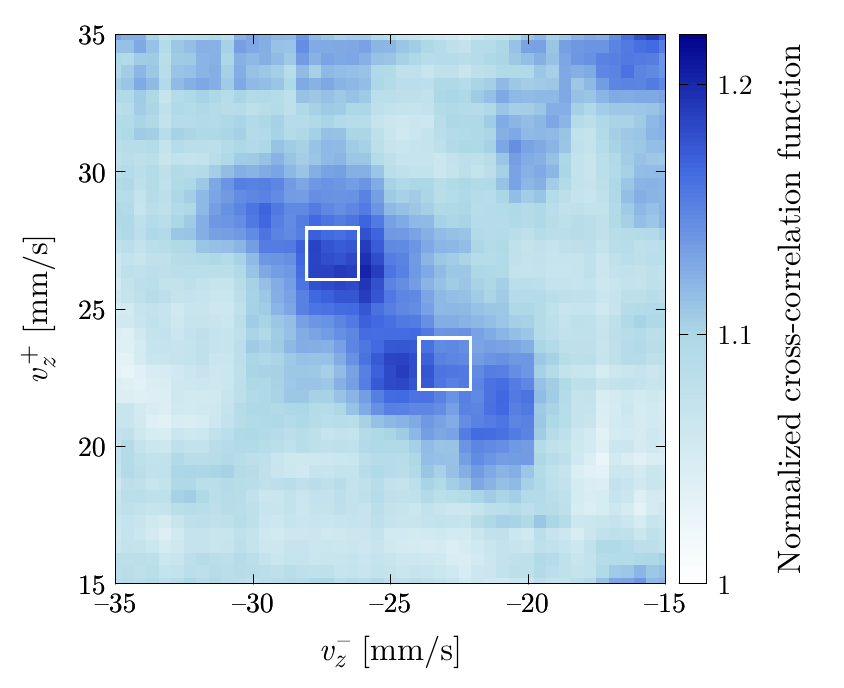} 
	\caption{ \label{Fig:3}
		Normalized cross-correlation $g^{(2)}(v_z^+, v_z^-)$. The velocities are measured along the $z$-axis and relative to the center-of-mass velocity of the atom pairs. A sliding average was performed to reduce the statistical noise.
		The correlation peak is elongated along the anti-diagonal because the source can emit in  several pairs of modes. The width of the correlation peak along the diagonal corresponds to the diffraction limit imposed by the spatial extent of the source.
		The white squares show the size and position in the plane $(v_z^+,v_z^-)$ of the integration volumes used to obtain the points in Fig.~\ref{Fig:4} for the set of modes 1.
	}
\end{figure}

The pairwise emission process is characterized by the normalized cross-correlation:
\begin{equation}
	\label{g2}
	g^{(2)}(v_z^+, v_z^-) = \frac{\langle n(v_z^+)\,n(v_z^-) \rangle}{\langle n(v_z^+) \rangle \langle n(v_z^-) \rangle} \; ,
\end{equation}
where $n(v_z^\pm)$ represents the number of atoms with a velocity $v_z^+ > 0$, or $v_z^- < 0$, along the $z$-axis and 0 along the $x$- and $y$-axes. Experimentally, we measure this correlation by counting the number of detected atoms inside two small integration volumes in velocity-space \cite{Supplemental}, and averaging their product over many realizations (as denoted by $\langle \cdot \rangle$). The correlation obtained in the experiment is displayed in Fig.~\ref{Fig:3}. A two-particle correlation centered around $v_z^+ = -v_z^- \simeq \SI{25}{\mm\per\s}$ is clearly visible, and confirms that atoms are indeed emitted in pairs with opposite velocities. Because the pair emission fulfills the quasi-momentum conservation strictly, but the energy conservation only loosely \cite{Bonneau2013}, our source emits several pairs of modes, as shown by the correlation peak which is elongated along the line $v_z^+ = -v_z^-$ \cite{Supplemental}.

If the pair production process is coherent, emitted pairs will be in a superposition of several pair states, each with well defined velocities. In other words, our source of atom pairs should produce pairs of entangled atoms. By filtering the velocities at the detector according to: $m v_z^+ = p$ or $p'$, and $mv_z^- = -p$ or $-p'$, where $m$ is the mass of the atom, we therefore expect to obtain a Bell state of the form (\ref{Eq:Psi}), expressed in the center-of-mass reference frame of the pairs.
The next step is to observe an interference between the two components of the superposition state with the interferometer in Fig.~\ref{Fig:1}. 
This is realized using Bragg diffraction of the atoms on a second
optical lattice oriented along the $z$-axis, distinct from the lattice driving the pair emission.
This Bragg lattice is pulsed first for \SI{100}{\micro\s} to realize the Bragg deflector ($\pi$-pulse), and then for \SI{50}{\micro\s} to realize the Bragg splitters ($\pi/2$-pulse).
During the whole time, the frequency difference between the laser beams forming the lattice is chirped to compensate for the atoms' free fall. 
The Bragg resonance is met when $v_z^\pm = \SI{\pm 25}{\mm\per\s}$ but the finite pulse duration broadens the Bragg energy condition such that all mode pairs $(p,-p')$, or $(-p,p')$, produced in the experiment are coupled with almost the same strength if they fulfill the Bragg momentum condition
\begin{equation}
	\label{eq:Bragg_condition}
	p+p' = \hbar k_\ell \; ,
\end{equation}
where $k_\ell = m/\hbar \times \SI{50}{\mm\per\s}$ is the lattice reciprocal vector.
This has two practical consequences. First, a single Bragg lattice simultaneously realizes the deflection, or the mixing, of the two pairs of modes $(p,-p')$ and $(-p,p')$, in contrast with the configuration shown in Fig.~\ref{Fig:1}, where two independent splitters are shown.
Second, since by construction the interferometer is closed for any pair of modes satisfying Eq.~(\ref{eq:Bragg_condition}), the same sequence of two successive Bragg lattices realizes several interferometers simultaneously.

We apply the deflecting pulse right after the transfer to the $m_J=0$ state, at $t_1 = \SI{1100}{\micro\s}$, where the time origin is set at the instant when the optical lattice driving the pair emission is switched on, and $t_1$ is the beginning of the pulse. To close the interferometer, the time $t_2$ for the splitting pulse is determined experimentally. This is achieved by performing a Hong--Ou--Mandel experiment \cite{Lopes2015}; that is, we vary the time at which the splitting pulse is applied and measure the probability of joint detection at velocities $v_z^\pm = \pm \SI{25}{\mm\per\s}$ (dashed lines in Fig.~\ref{Fig:1}). The interferometer is closed at the Hong--Ou--Mandel dip, that is when the joint detection probability is minimum. In our experiment, this occurs when the Bragg splitting pulse starts at $t_2 = \SI{1950}{\micro\s}$ \cite{Supplemental}.

Ideally, one would vary the phase difference $(\phi_A-\phi_B)$ in a controlled manner to observe the modulation predicted in Eq.~(\ref{Eq:correl}). This is not possible with the setup described here because the two splitters are realized with a single Bragg lattice.
Active control of the phase difference could be achieved using independent Bragg lattices for the splitters $A$ and $B$, and we intend to implement this procedure in the future.
However, we still have a way to probe different relative phases in the current setup by filtering modes for which the Bragg energy condition is not exactly satisfied, which adds a velocity-dependent contribution to $(\phi_A-\phi_B)$ \cite{Supplemental}. We therefore obtain different relative phases by filtering different output momenta.
 
\begin{figure}
	\includegraphics{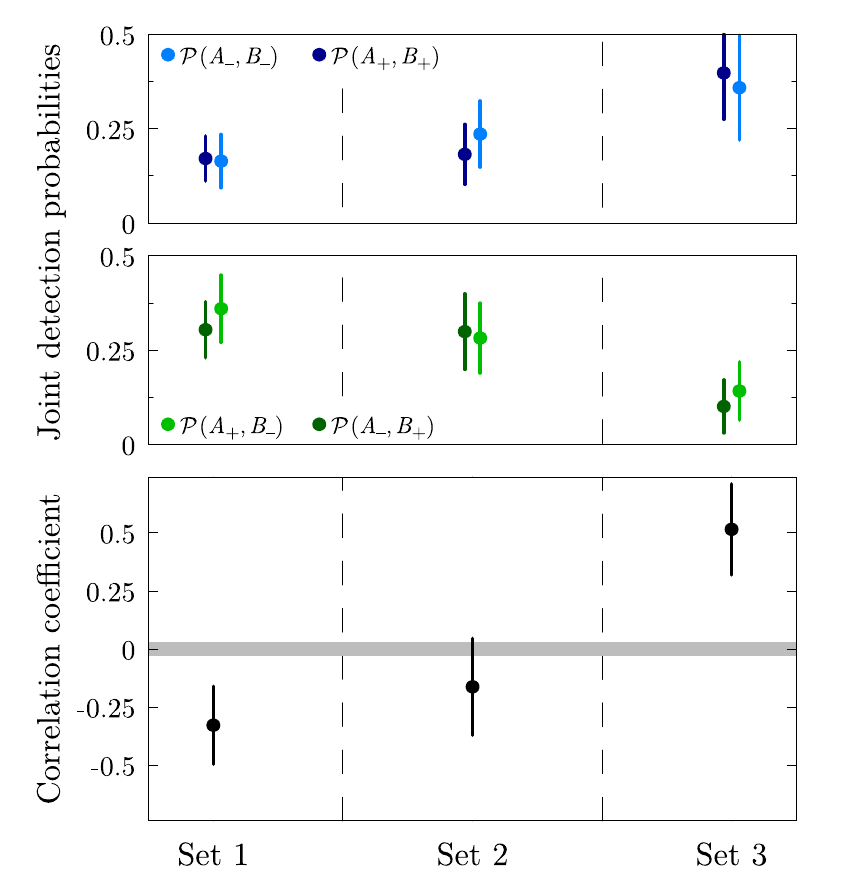}
	\caption{ \label{Fig:4}
		Joint detection probabilities measured at the output of the four-mode interferometer for three independent sets of momentum modes $(p, -p')$, and $(-p, p')$.
		The lower graph displays the correlation coefficient, $E$. The gray line represents the zero level of this coefficient, calibrated using different combinations of the same modes for which no two-particle interference can occur by construction; the width of the line is the uncertainty on the zero level.
		The velocities $v_z^+$ corresponding to the modes $p$ are \SIlist{27;29;31}{\mm\per\s} for sets 1, 2 and 3, respectively. The velocities corresponding to $p'$ can be deduced from Eq.~(\ref{eq:Bragg_condition}).
		Averages were taken over \num{2218} repetitions of the experiment.
		Error bars denote the statistical uncertainty and are obtained by bootstrapping.
	}
\end{figure}

In our experiment, the multiplexed character of the interferometer allows us to select three sets of mode pairs $(p, -p')$, and $(-p,p')$, for which the interferometer is closed with different relative phases $(\phi_A-\phi_B)$. A simple model for the Bragg diffraction \cite{Supplemental} indicates that the relative phases for these three sets span an interval of about \SI{100}{\degree}. For each set we measure the joint detection probabilities in the output modes using small integration volumes in velocity space \cite{Supplemental}. The white squares in Fig.~\ref{Fig:3} show the corresponding areas in the $(v_z^+, v_z^-)$ plane for one of the sets. The size of the integration volumes is a compromise between two opposing constraints: maximizing the signal-to-noise ratio and minimizing the variations across the volume of the phase imprinted upon diffraction. With our settings \cite{Supplemental}, the average population in one integration volume is \num{0.2} atoms per repetition (corrected for the \SI{25}{\percent} detection efficiency) and the phase varies by up to \SI{50}{\degree}.

Figure~\ref{Fig:4} displays the result of our measurements on each set.
The upper two graphs show the four joint detection probabilities. As expected from Eqs.~(\ref{Eq:PjointSym}) and (\ref{Eq:PjointAsym}), the values of $\mathcal P (A_+,B_+)$ and $\mathcal P (A_-,B_-)$ on the one hand, and $\mathcal P (A_+,B_-)$ and $\mathcal P (A_-,B_+)$ on the other, appear to be correlated. Note that, for each set, the sum of all four joint detection probabilities is equal to unity by construction.
The lower graph shows the correlation coefficient $E$ defined in Eq.~(\ref{Eq:correl}). We observe that, for at least one set of modes, this coefficient takes a non-zero value (set 3 gives $E=\num{0.51 \pm 0.20}$). 
We have also used our data to verify the zero level of $E$: By combining the modes analyzed in Fig.~\ref{Fig:4} in a way that avoids two-particle interferences by construction, we can build 18 sets of modes that should exhibit a zero correlation coefficient \cite{Supplemental}. For those reference sets, we find indeed $E=\num{0.00}$ with a statistical uncertainty of \num{0.03} (gray line in the lower graph of Fig.~\ref{Fig:4}).

Our results thus rule out the possibility of a completely mixed state at the input of the interferometer.
To make a claim about the presence of entanglement, we would need to observe the modulation of $E$ when we vary the phase difference $(\phi_A-\phi_B)$.
This is best achieved by introducing separate Bragg splitters, and performing a correlation measurement on a single set of momentum modes to render common any velocity dependent phase.
A contrast of the oscillation in excess of $1/\sqrt 2$ would permit the observation of a Bell inequality violation for freely falling massive particles using their momentum degree of freedom.
Finally, we note that the setup described here can in principle be adapted to mix the mode $p$ with $p'$, and $-p$ with $-p'$, by changing the reciprocal wavevector of the Bragg lattices. This variant, where the trajectories of the two atoms never cross, can also lead to a violation of a Bell inequality, in a situation where non-locality is more striking.

\begin{acknowledgments}
	The research leading to these results has received funding from the European Research Council and from the People Programme (Marie Curie Actions) under the European Union's Seventh Framework Programme (FP7/2007-2013) and H2020 Programme (2014-2020) / ERC grant agreement $\text{n}^\circ$267775, and REA grant agreements $\text{n}^\circ$618760 and 704832.
	We also acknowledge funding from the ANR through the grant agreement $\text{n}^\circ$15-CE30-0017 and support from Churchill College, Cambridge.
\end{acknowledgments}


\section{Appendix}

\subsection{Optical lattice for pair emission}

The optical lattice driving the dynamical instability at the origin of the pair creation has a period $a = \SI{550}{\nano\m}$ and a depth of \SI{0.45}{\Erec}, where $\si{\Erec} = \pi^2 \hbar^2 / 2ma^2$ is the recoil energy and $m$ is the mass of an atom. The frequency difference between the two laser beams forming the lattice is $\nu = \SI{105}{\kilo\Hz}$, resulting in a velocity $\nu a = \SI{57}{\mm\per\s}$ for the motion of the standing wave in the laboratory frame of reference.

\subsection{Integration volumes for counting the atom numbers}

Depending on the observable, we choose different integration volumes in velocity space in order to optimize the signal-to-noise ratio. In Tab.~\ref{Tab:Volumes}, we summarize the integration volumes used to count the number of atoms for each graph of the main text and supplemental material.
{
\setlength{\tabcolsep}{5pt}
\ctable[
	caption = {Integration volumes. Rectangular boxes have a size $\delta v_x$, $\delta v_y$ and $\delta v_z$ along $x$, $y$ and $z$, respectively. Cylindrical boxes are oriented along $z$; their diameter is $\delta v_x = \delta v_y$ and their length is $\delta v_z$. All sizes are given in \si{\mm\per\s}.},
	label = Tab:Volumes,
	mincapwidth = \columnwidth,
	pos = h]
{lllll}{}
{
																					\FL
		& box shape		& $\delta v_x$		& $\delta v_y$		& $\delta v_z$		\ML
Fig.~2	& rectangular	& \tablenum{9.2}	& \tablenum{2.4}	& \tablenum{0.9}	\NN
Fig.~3	& cylindrical	& \tablenum{32.2}	& \tablenum{32.2}	& \tablenum{2.8}	\NN
Fig.~4	& cylindrical	& \tablenum{4.0}		& \tablenum{4.0}	& \tablenum{2.0}	\NN
Fig.~S1 (left) 	& cylindrical & \tablenum{18.4}	& \tablenum{18.4}	& \tablenum{1.8}	\NN
Fig.~S1 (right)	& cylindrical & \tablenum{18.4}	& \tablenum{18.4}	& \tablenum{0.9}	\NN
Fig.~S3	& cylindrical	& \tablenum{4.0}	& \tablenum{4.0}	& \tablenum{2.6}	\LL
}
}

\subsection{Normalized cross-correlation}

\begin{figure}
	\includegraphics{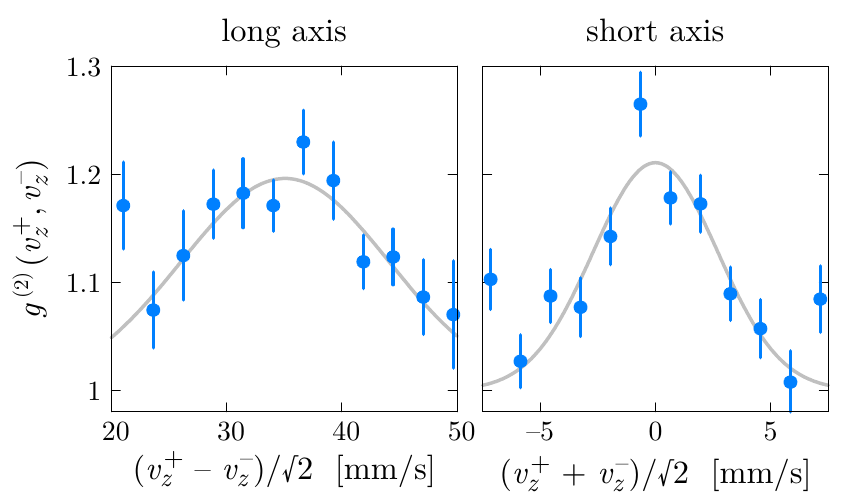}
	\caption{ \label{Fig:S1}
		Projections of the two-dimensional cross-correlation function on its long and short axes. The blue points represent the experimental data. The error bars represent the statistical uncertainty and are obtained by bootstrapping. The grey lines are Gaussian fits with an offset fixed at unity.
	}
\end{figure}

The normalized cross-correlation $g^{(2)}(v_z^+, v_z^-)$ shown in Fig.~3 of the main text displays a peak centered around $v_z^+ = -v_z^- \simeq \SI{25}{\mm\per\s}$. This peak is elongated along the line $v_z^+ = -v_z^-$, indicating that the source emits several pairs of modes.
Projections of the two-dimensional cross-correlation function along the lines $v_z^+ = -v_z^-$ and $v_z^+ - v_z^- = \SI{50}{\mm\per\s}$, corresponding to the long and short axes of the correlation peaks, respectively, are given in Fig.~\ref{Fig:S1}.
Unlike the two-dimensional map displayed in Fig.~3 of the main text, no sliding average was performed and all experimental points are statistically independent.
The different amplitudes of the correlation peak along the long and short axes stem from the different integration volumes.
A Gaussian fit yields the half-widths (standard deviation) $\sigma = \SI{9.0 \pm 2.3}{\mm\per\s}$ for the long axis, and $\sigma = \SI{2.7 \pm 0.7}{\mm\per\s}$ for the short axis. These values are to be compared to the half-width of the auto-correlation functions, $\sigma_\text{auto} = \SI{1.9 \pm 0.4}{\mm\per\s}$, which is the diffraction limit of our source.

\subsection{Timing of the Bragg pulses}

\begin{figure}
	\includegraphics{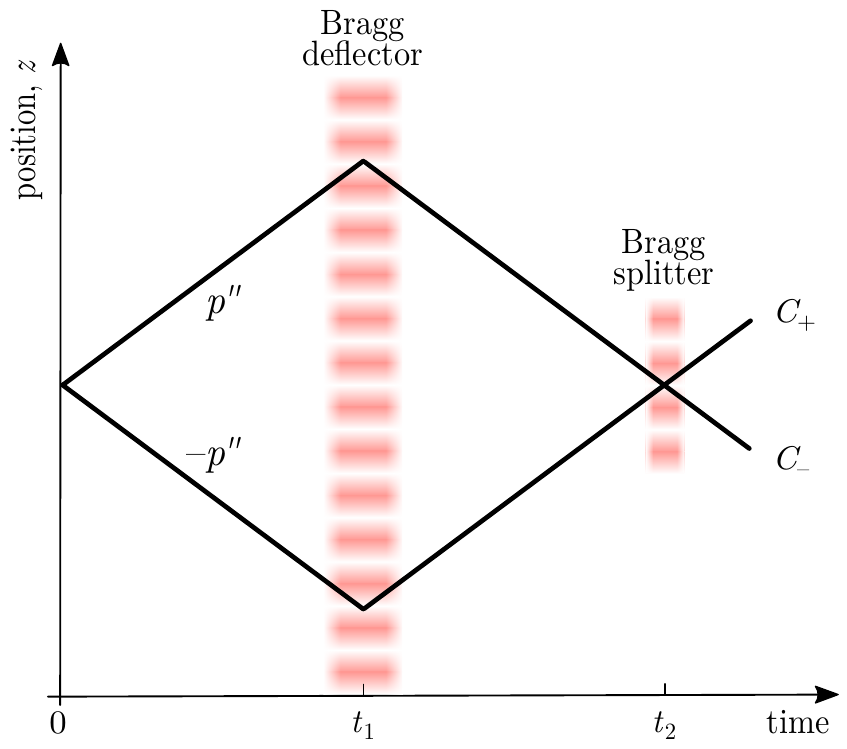}
	\caption{
		\label{Fig:S2}
		Diagram of the Hong--Ou--Mandel interferometer. By filtering only two output modes corresponding to the initial pair state $|p'',-p''\rangle$, the four-mode interferometer folds onto a two-mode interferometer. The Hong--Ou--Mandel effect occurs when the Bragg splitting pulse mixes the two input modes $p''$ and $-p''$. It manifests as a reduction of the probability of joint detection in the output modes $C_+$ and $C_-$, shown in Fig.~\ref{Fig:S3}.
	 }
\end{figure}

In principle, the interferometer is closed when the time $t_2$ at which the mixing is realized equals twice the time $t_1$ at which the deflection is realized. However, neither the pair emission, nor the Bragg diffraction, occur at a well defined time and we have to determine experimentally the time at which the Bragg splitting pulse must be applied in order to close the interferometer.
We solve this problem by performing a Hong--Ou--Mandel experiment. This is achieved by filtering two symmetric output modes, $C_+$ and $C_-$, associated with the input state $|p'',-p''\rangle$, where $2\,p'' = \hbar k_\ell$ (see Fig.~\ref{Fig:S2}).
In the experiment, these modes correspond to the velocities $v_z^+ = -v_z^- = \SI{25}{\mm\per\s}$, which are located at the maxima of the initial velocity distribution of the emitted atom pairs. We then vary the time at which the Bragg splitting pulse is applied, and measure the probability of joint detection in the two output modes:
\begin{equation}
	\mathcal P(C_+,C_-) = 2 \Lambda^{-1} \langle n(p'') \, n(-p'') \rangle \; ,
\end{equation}
where the normalization factor is given by
\begin{multline}
	\Lambda = \langle n(p'')(n(p'')-1) \rangle + \langle n(-p'')(n(-p'')-1) \rangle \\
		+ 2 \langle n(p'') n(-p'') \rangle \; ,
\end{multline}
and we have used the notation $n(\pm p'')$ instead of $n(v_z^\pm)$, with $mv_z^\pm = \pm p''$.

\begin{figure}
	\includegraphics{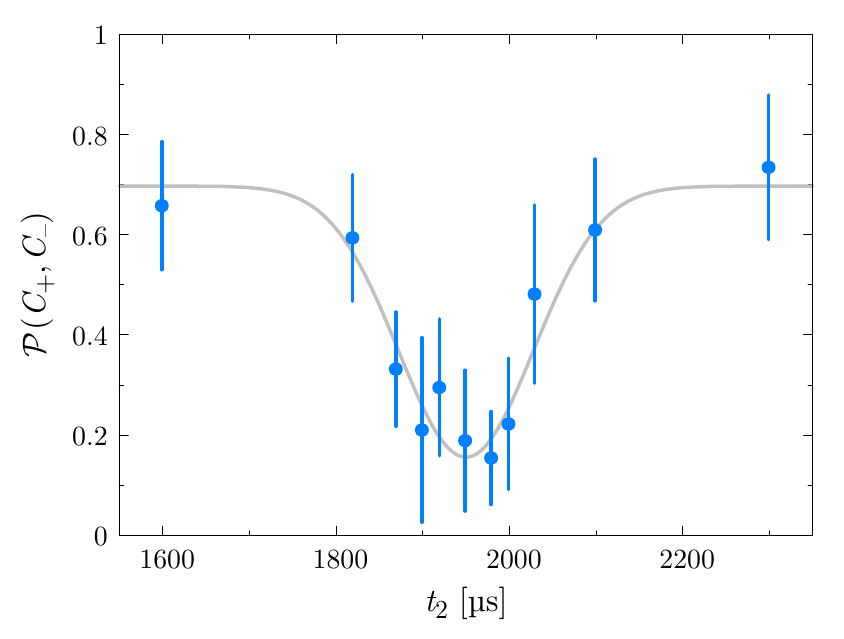}
	\caption{
		\label{Fig:S3}
		Joint detection probability in the two symmetric output modes as a function of the time at which the Bragg splitting pulse is applied. The blue points represent the experimental data. The error bars represent the statistical uncertainty and are obtained by bootstrapping. The grey line is a Gaussian fit. The reduction of the joint detection probability at $t_2 = \SI{1950}{\micro\s}$ results from the Hong--Ou--Mandel effect, and signals that the interferometer is closed.
	 }
\end{figure}

In a closed interferometer, the \enquote{which-path} information is erased and the two atoms of a pair become indistinguishable after the Bragg splitter. A two-particle interference then results in the cancellation of the joint detection probability for bosons. We show the result of this measurement in Fig.~\ref{Fig:S3}. The dip in the joint detection probability is clearly visible when the Bragg splitting pulse begins at time $t_2 = \SI{1950}{\micro\s}$, and we use this timing to realize the four-mode interferometer.

\section{Bragg diffraction model}

The Bragg reflectors and splitters are realized by Bragg diffraction on a vertical, moving optical lattice formed by the interference pattern of two laser beams with slightly different frequencies. The frequency difference between the two beams forming the lattice is chirped to compensate for the atoms' free fall.
In the limit of a shallow lattice, i.e. when the lattice depth is smaller than the recoil energy $E_\text{rec} = \hbar^2 k_\ell^2/2m$, Bragg diffraction couples only pairs of momentum states $(p,-p')$, or $(-p,p')$, satisfying both momentum conservation, $p+p' = \hbar k_\ell$, and energy conservation, $p^2/2m = p'^2/2m$. If the interaction time between the atoms and the lattice is short, however, the energy conservation condition is not strict. In contrast, the momentum conservation condition will always be strictly fulfilled because of the spatial extension of the optical lattice and the high velocity resolution of our detector.

\subsubsection{Diffraction at the Bragg energy condition}

We consider here the pair of input modes $p$ and $-p'$ fulfilling exactly the Bragg energy condition. We write the coupling Hamiltonian in the basis $\{|p\rangle,|-p'\rangle\}$ as:
\begin{equation}
	\hat H = \frac{\hbar\Omega}{2}
	\left(
		\begin{matrix}
			0 & e^{i\phi} \\ e^{-i\phi} & 0
		\end{matrix}
	\right) \, ,
\end{equation}
where $\Omega/2\pi$ is the two-photon Rabi frequency and $\phi$ is the phase difference between the two laser beams forming the Bragg lattice.
The Bragg lattice drives a Rabi oscillation between the two modes $p$ and $-p'$. The evolution operator describing this dynamics takes the simple form:
\begin{equation}
	\hat U(t) \equiv e^{-i \hat H t/\hbar} = 
	\left(
		\begin{matrix}
			\cos\left(\Omega t/2\right) & -i\,e^{-i\phi} \sin\left(\Omega t/2\right) \\
			-i\,e^{i\phi} \sin\left(\Omega t/2\right) & \cos\left(\Omega t/2\right)
		\end{matrix}
	\right) \, ,
\end{equation}
where the time origin is set at the instant when the laser beams are switched on.
An interaction time $t = \pi/\Omega$ ($\pi$-pulse) turns an input state $|p\rangle$ into an output state $|-p'\rangle$, and an input state $|-p'\rangle$ into an output state $|p\rangle$; it therefore realizes a Bragg deflector. Similarly, an interaction time $t = \pi/2\Omega$ ($\pi/2$-pulse) turns $|p\rangle$ or $|-p'\rangle$ into a superposition with equal weights of $|p\rangle$ and $|-p'\rangle$; it therefore realizes a {50:50} Bragg splitter.

In our interferometer, a $\pi$-pulse and a $\pi/2$-pulse are successively applied to realize the deflection and the splitting. Using the subscripts $D$ and $A$ to label the deflecting pulse and the splitting pulse $A$, respectively, and omitting overall phase factors due to propagation, we therefore obtain the output modes $A_+$ and $A_-$ by writing:
\begin{align}
	\left(\begin{matrix} A_+ \\ A_- \end{matrix}\right)
	& = \hat U_A(\pi/2\Omega)\,\hat U_D(\pi/\Omega)\,
		\left(\begin{matrix} p \\ -p' \end{matrix}\right) \\
	& = \mfrac{-1}{\sqrt 2}
		\left(
		\begin{matrix}
			e^{-i(\phi_A-\phi_D)} & i\,e^{-i\phi_D} \\
			i\,e^{i\phi_D} & e^{i(\phi_A-\phi_D)}
		\end{matrix}
		\right)
		\left(\begin{matrix} p \\ -p' \end{matrix}\right) \, .
\end{align}
The same reasoning applies if we consider the pair of input modes $p'$ and $-p$. We then obtain:
\begin{align}
	\left(\begin{matrix} B_+ \\ B_- \end{matrix}\right)
	& = \hat U_B(\pi/2\Omega)\,\hat U_D(\pi/\Omega)\,
		\left(\begin{matrix} p' \\ -p \end{matrix}\right) \\
	& = \mfrac{-1}{\sqrt 2}
		\left(
		\begin{matrix}
			e^{-i(\phi_B-\phi_D)} & i\,e^{-i\phi_D} \\
			i\,e^{i\phi_D} & e^{i(\phi_B-\phi_D)}
		\end{matrix}
		\right)
		\left(\begin{matrix} p' \\ -p \end{matrix}\right) \, .
\end{align}
Equations~(S4–S7) directly give Eqs.~(2–5) in the main text.

\subsubsection{Diffraction away from the Bragg energy condition}

We now consider a pair of input modes $p$ and $-p'$ for which the Bragg energy condition is not exactly satisfied, meaning that the input and output states have slightly different energies.
We introduce the detuning $\hbar\delta = p^2/2m - p'^2/2m$ and assume $\delta > 0$. To first order in $\delta/\Omega$, the evolution operator in the basis $\{|p\rangle,|-p'\rangle\}$ is modified according to:
\begin{equation}
	\hat U(t) \simeq
	\left(
		\begin{matrix}
			e^{-i\delta t/2} \cos\left(\Omega t/2\right) &
			-i\,e^{-i(\phi+\delta t/2)} \sin\left(\Omega t/2\right) \\
			-i\,e^{i(\phi+\delta t/2)} \sin\left(\Omega t/2\right) &
			e^{i\delta t/2} \cos\left(\Omega t/2\right)
		\end{matrix}
	\right) \, .
\end{equation}
If we consider instead the input states $p'$ and $-p$, but keep the same definition for $\delta$, we must take care to replace $\delta$ by $-\delta$ in this evolution operator.
Compared to the resonant case, one sees that an additional phase $\delta t$ is accumulated between the components $|p\rangle$ and $|-p'\rangle$ during the interaction with the Bragg lattice. At the output of the interferometer, the modes $A_\pm$ and $B_\pm$ are now given by the matrix equations
\begin{align}
	\label{Eq:MatrixA}
	\left(\begin{matrix} A_+ \\ A_- \end{matrix}\right)
	& \simeq \mfrac{-1}{\sqrt 2}
		\left(
		\begin{matrix}
			e^{-i(\phi_A-\phi_D-\pi\delta/4\Omega)} & i\,e^{-i(\phi_D+3\pi\delta/4\Omega)} \\
			i\,e^{i(\phi_D+3\pi\delta/4\Omega)} & e^{i(\phi_A-\phi_D-\pi\delta/4\Omega)}
		\end{matrix}
		\right)
		\left(\begin{matrix} p \\ -p' \end{matrix}\right) \\
	\intertext{and}
	\label{Eq:MatrixB}
	\left(\begin{matrix} B_+ \\ B_- \end{matrix}\right)
	& \simeq \mfrac{-1}{\sqrt 2}
		\left(
		\begin{matrix}
			e^{-i(\phi_B-\phi_D+\pi\delta/4\Omega)} & i\,e^{-i(\phi_D-3\pi\delta/4\Omega)} \\
			i\,e^{i(\phi_D-3\pi\delta/4\Omega)} & e^{i(\phi_B-\phi_D+\pi\delta/4\Omega)}
		\end{matrix}
		\right)
		\left(\begin{matrix} p' \\ -p \end{matrix}\right) \, .
\end{align}

Inverting the matrix equations (\ref{Eq:MatrixA}) and (\ref{Eq:MatrixB}), we can express the entangled state $|\psi\rangle = \frac{1}{\sqrt{2}} \left( |p,-p\rangle + |p',-p'\rangle \right)$ at the output of the interferometer as
\begin{equation}
	\begin{split}
	|\Psi_\text{out}\rangle \simeq
		\smash{\mfrac{1}{2\sqrt 2}} \Big[ \; 
			- & i \big( e^{i(\phi_A-\pi\delta/\Omega)} + e^{i(\phi_B+\pi\delta/\Omega)} \big) |A_+,B_+\rangle \\
			+ & \big( e^{i(\phi_A-\phi_B-\pi\delta/2\Omega)} - e^{3i\pi\delta/2\Omega} \big) |A_+,B_-\rangle \\
			+ & \big( e^{-i(\phi_A-\phi_B-\pi\delta/2\Omega)} - e^{-3i\pi\delta/2\Omega} \big) |A_-,B_+\rangle \\
			- & i \big( e^{-i(\phi_A-\pi\delta/\Omega)} + e^{-i(\phi_B+\pi\delta/\Omega)} \big) |A_-,B_-\rangle
		\; \Big] \; .
	\end{split}
\end{equation}
We finally obtain the joint detection probabilities
\begin{align}
	\mathcal P (A_\pm,B_\pm) & \simeq \mfrac{1}{2} \cos^2 \big[ (\phi_A-\phi_B-2\pi\delta/\Omega)/2 \big] \; , \\
	\mathcal P (A_\pm,B_\mp) & \simeq \mfrac{1}{2} \sin^2 \big[ (\phi_A-\phi_B-2\pi\delta/\Omega)/2 \big] \; .
\end{align}
To first order in $\delta/\Omega$, the mismatch in the Bragg energy condition thus adds an offset $-2\pi\delta/\Omega$ to the relative phase $(\phi_A - \phi_B)$. This off-resonance contribution depends on $p$ and $p'$ through the detuning $\delta$. 

In the sets of modes 1, 2 and 3 shown in Fig.~4, the mode $p$ corresponds, respectively, to the velocities $v_z^+ = \text{\SIlist{27.0;29.1;31.1}{\mm\per\s}}$, and the mode $p'$ corresponds to the velocities $v_z^+ = \text{\SIlist{23.0;20.9;18.9}{\mm\per\s}}$.
The mismatch in the Bragg energy condition for these three sets of modes are thus: $\delta_1/2\pi = \SI{0.9}{\kilo\hertz}$, $\delta_2/2\pi = \SI{1.9}{\kilo\hertz}$ and $\delta_3/2\pi = \SI{2.9}{\kilo\hertz}$. For these values of the detuning $\delta$, the condition $\delta \ll \Omega$ is only marginally satisfied and the lowest order approximation overestimates the relative phases by about \SI{30}{\percent}. For a better estimation, we wrote the exact evolution operator for the two-mode dynamics, and numerically calculated the additional phases with respect to the resonant case. We found \text{\SIlist{-43;-94;-144}{\degree}} for sets 1, 2 and 3, respectively.

\subsection{Experimental measurement of the joint detection probabilities}

The probabilities of joint detection in the output modes $A_\pm$ and $B_\pm$ are measured by counting the number of atoms with velocities $mv_z^+ = p$ or $p'$, and $mv_z^- = -p$ or $-p'$, and using the relations
\begin{align}
	\mathcal P(A_+,B_+) & = \Lambda^{-1} \langle n(p) \, n(p') \rangle \; , \\
	\mathcal P(A_-,B_-) & = \Lambda^{-1} \langle n(-p) \, n(-p') \rangle \; , \\
	\mathcal P(A_+,B_-) & = \Lambda^{-1} \langle n(p) \, n(-p) \rangle \; , \\
	\mathcal P(A_-,B_+) & = \Lambda^{-1} \langle n(-p') \, n(p') \rangle \; ,
\end{align}
where the normalization factor is given by
\begin{multline}
	\Lambda = \langle n(p) n(p') \rangle + \langle n(-p) n(-p') \rangle \\
		+ \langle n(p) n(-p) \rangle + \langle n(-p') n(p') \rangle \; ,
\end{multline}
and we have used the notation $n(\pm p)$ or $n(\pm p')$ instead of $n(v_z^\pm)$, with $mv_z^\pm = \pm p$ or $\pm p'$.

\subsection{Zero level of the correlation coefficient}

\begin{figure}
	\includegraphics{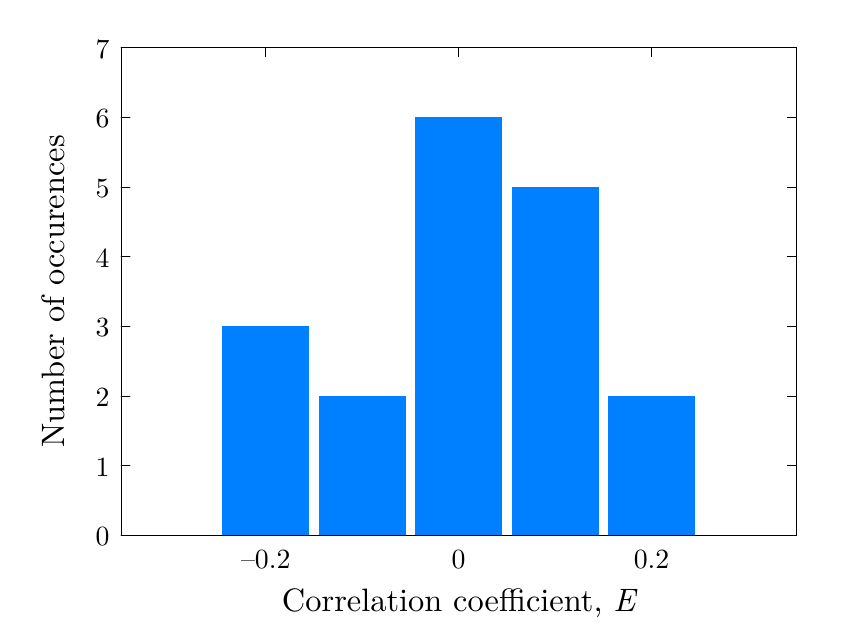}
	\caption{ \label{Fig:S4}
		Histogram of the correlation coefficient $E$ measured with 18 combinations of the modes analyzed in Fig.~4 of the main text for which no two-particle interference can occur by construction. The distribution is peaked around zero, showing that there is no bias in the evaluation of the correlation coefficient.
	 }
\end{figure}
In order to confirm the zero level of the correlation coefficient $E$, we have constructed a correlation coefficient using 18 combinations of the modes analyzed in Fig.~4 of the main text for which no two-particle interference can occur by construction. Denoting by $A_\pm^{(i)}$ and $B_\pm^{(i)}$ the output modes of the set $i$ ($i=1, 2, 3$), those combinations are of the form $\{A_+^{(i)}, A_-^{(j)}, B_+^{(k)}, B_-^{(l)}\}$ with $i \neq k,l$ and $j \neq k, l$.
The mean values of the joint detection probabilities measured with these reference sets are all close to 1/4, as summarized in Tab.~\ref{Tab:Proba}. In Fig.~\ref{Fig:S4}, we show a histogram of the corresponding values of $E$. The distribution has a mean value of \num{-0.001} and a standard deviation of \num{0.125}. The uncertainty on the mean value is $\num{0.125}/\sqrt{18} = \num{0.029}$. These calibration measurements give us confidence that we have no systematic bias in the estimation of the correlation coefficient.
{
\smallskip
\setlength{\tabcolsep}{5pt}
\ctable[
	caption = {Mean values of the joint detection probabilities for uncorrelated data.},
	label = Tab:Proba,
	mincapwidth = 0.75\columnwidth,
	pos = H]
{ll}{}
{
												\FL
$\mathcal P (A_+,B_+)$	& \num{0.258 \pm 0.015}	\NN
$\mathcal P (A_-,B_-)$	& \num{0.241 \pm 0.011}	\NN
$\mathcal P (A_+,B_-)$	& \num{0.259 \pm 0.018}	\NN
$\mathcal P (A_-,B_+)$	& \num{0.242 \pm 0.013}	\LL
}
}

\bibliographystyle{apsrev4-1}
\bibliography{2p4m-v24_arXiv}

\begin{thebibliography}{29}%
\makeatletter
\providecommand \@ifxundefined [1]{%
 \@ifx{#1\undefined}
}%
\providecommand \@ifnum [1]{%
 \ifnum #1\expandafter \@firstoftwo
 \else \expandafter \@secondoftwo
 \fi
}%
\providecommand \@ifx [1]{%
 \ifx #1\expandafter \@firstoftwo
 \else \expandafter \@secondoftwo
 \fi
}%
\providecommand \natexlab [1]{#1}%
\providecommand \enquote  [1]{``#1''}%
\providecommand \bibnamefont  [1]{#1}%
\providecommand \bibfnamefont [1]{#1}%
\providecommand \citenamefont [1]{#1}%
\providecommand \href@noop [0]{\@secondoftwo}%
\providecommand \href [0]{\begingroup \@sanitize@url \@href}%
\providecommand \@href[1]{\@@startlink{#1}\@@href}%
\providecommand \@@href[1]{\endgroup#1\@@endlink}%
\providecommand \@sanitize@url [0]{\catcode `\\12\catcode `\$12\catcode
  `\&12\catcode `\#12\catcode `\^12\catcode `\_12\catcode `\%12\relax}%
\providecommand \@@startlink[1]{}%
\providecommand \@@endlink[0]{}%
\providecommand \url  [0]{\begingroup\@sanitize@url \@url }%
\providecommand \@url [1]{\endgroup\@href {#1}{\urlprefix }}%
\providecommand \urlprefix  [0]{URL }%
\providecommand \Eprint [0]{\href }%
\providecommand \doibase [0]{http://dx.doi.org/}%
\providecommand \selectlanguage [0]{\@gobble}%
\providecommand \bibinfo  [0]{\@secondoftwo}%
\providecommand \bibfield  [0]{\@secondoftwo}%
\providecommand \translation [1]{[#1]}%
\providecommand \BibitemOpen [0]{}%
\providecommand \bibitemStop [0]{}%
\providecommand \bibitemNoStop [0]{.\EOS\space}%
\providecommand \EOS [0]{\spacefactor3000\relax}%
\providecommand \BibitemShut  [1]{\csname bibitem#1\endcsname}%
\let\auto@bib@innerbib\@empty
\bibitem [{\citenamefont {Dowling}\ and\ \citenamefont
  {Milburn}(2003)}]{dowling2003quantum}%
  \BibitemOpen
  \bibfield  {author} {\bibinfo {author} {\bibfnamefont {J.~P.}\ \bibnamefont
  {Dowling}}\ and\ \bibinfo {author} {\bibfnamefont {G.~J.}\ \bibnamefont
  {Milburn}},\ }\href@noop {} {\bibfield  {journal} {\bibinfo  {journal} {Phil.
  Trans. R. Soc. Lond. A}\ }\textbf {\bibinfo {volume} {361}},\ \bibinfo
  {pages} {1655} (\bibinfo {year} {2003})}\BibitemShut {NoStop}%
\bibitem [{\citenamefont {Aspect}(2004)}]{Aspect:2004introductionsuqm}%
  \BibitemOpen
  \bibfield  {author} {\bibinfo {author} {\bibfnamefont {A.}~\bibnamefont
  {Aspect}},\ }\enquote {\bibinfo {title} {Introduction: John bell and the
  second quantum revolution},}\ in\ \href@noop {} {\emph {\bibinfo {booktitle}
  {Speakable and Unspeakable in Quantum Mechanics: Collected Papers on Quantum
  Philosophy}}}\ (\bibinfo  {publisher} {Cambridge University Press},\ \bibinfo
  {address} {Cambridge},\ \bibinfo {year} {2004})\ pp.\ \bibinfo {pages}
  {xvii--xl}\BibitemShut {NoStop}%
\bibitem [{\citenamefont {Feynman}(1982)}]{Feynman:1982yx}%
  \BibitemOpen
  \bibfield  {author} {\bibinfo {author} {\bibfnamefont {R.~P.}\ \bibnamefont
  {Feynman}},\ }\href@noop {} {\bibfield  {journal} {\bibinfo  {journal} {Int.
  J. Theor. Phys.}\ }\textbf {\bibinfo {volume} {21}},\ \bibinfo {pages} {467}
  (\bibinfo {year} {1982})}\BibitemShut {NoStop}%
\bibitem [{\citenamefont {Gisin}\ and\ \citenamefont
  {Rigo}(1995)}]{gisin1995relevant}%
  \BibitemOpen
  \bibfield  {author} {\bibinfo {author} {\bibfnamefont {N.}~\bibnamefont
  {Gisin}}\ and\ \bibinfo {author} {\bibfnamefont {M.}~\bibnamefont {Rigo}},\
  }\href@noop {} {\bibfield  {journal} {\bibinfo  {journal} {J. Phys. A: Math.
  Gen.}\ }\textbf {\bibinfo {volume} {28}},\ \bibinfo {pages} {7375} (\bibinfo
  {year} {1995})}\BibitemShut {NoStop}%
\bibitem [{\citenamefont {Aspect}(2002)}]{Aspect:2002quantumunsp}%
  \BibitemOpen
  \bibfield  {author} {\bibinfo {author} {\bibfnamefont {A.}~\bibnamefont
  {Aspect}},\ }\enquote {\bibinfo {title} {Bell's {Theorem}: {The} {Naive}
  {View} of an {Experimentalist}},}\ in\ \href@noop {} {\emph {\bibinfo
  {booktitle} {Quantum (Un)speakables: From Bell to Quantum Information}}}\
  (\bibinfo  {publisher} {Springer},\ \bibinfo {address} {Berlin, Heidelberg},\
  \bibinfo {year} {2002})\ pp.\ \bibinfo {pages} {119--153},\ \bibinfo {note}
  {available at \texttt{arXiv:quant-ph/0402001}}\BibitemShut {NoStop}%
\bibitem [{\citenamefont {Tichy}\ \emph {et~al.}(2011)\citenamefont {Tichy},
  \citenamefont {Mintert},\ and\ \citenamefont
  {Buchleitner}}]{tichy2011essential}%
  \BibitemOpen
  \bibfield  {author} {\bibinfo {author} {\bibfnamefont {M.~C.}\ \bibnamefont
  {Tichy}}, \bibinfo {author} {\bibfnamefont {F.}~\bibnamefont {Mintert}}, \
  and\ \bibinfo {author} {\bibfnamefont {A.}~\bibnamefont {Buchleitner}},\
  }\href@noop {} {\bibfield  {journal} {\bibinfo  {journal} {J. Phys. B: At.
  Mol. Opt.}\ }\textbf {\bibinfo {volume} {44}},\ \bibinfo {pages} {192001}
  (\bibinfo {year} {2011})}\BibitemShut {NoStop}%
\bibitem [{\citenamefont {Bell}(1964)}]{Bell:1964cr}%
  \BibitemOpen
  \bibfield  {author} {\bibinfo {author} {\bibfnamefont {J.~S.}\ \bibnamefont
  {Bell}},\ }\href@noop {} {\bibfield  {journal} {\bibinfo  {journal}
  {Physics}\ }\textbf {\bibinfo {volume} {1}},\ \bibinfo {pages} {195}
  (\bibinfo {year} {1964})}\BibitemShut {NoStop}%
\bibitem [{\citenamefont {Einstein}\ \emph {et~al.}(1935)\citenamefont
  {Einstein}, \citenamefont {Podolsky},\ and\ \citenamefont
  {Rosen}}]{Einstein1935}%
  \BibitemOpen
  \bibfield  {author} {\bibinfo {author} {\bibfnamefont {A.}~\bibnamefont
  {Einstein}}, \bibinfo {author} {\bibfnamefont {B.}~\bibnamefont {Podolsky}},
  \ and\ \bibinfo {author} {\bibfnamefont {N.}~\bibnamefont {Rosen}},\ }\href
  {\doibase 10.1103/PhysRev.47.777} {\bibfield  {journal} {\bibinfo  {journal}
  {Phys. Rev.}\ }\textbf {\bibinfo {volume} {47}},\ \bibinfo {pages} {777}
  (\bibinfo {year} {1935})}\BibitemShut {NoStop}%
\bibitem [{\citenamefont {Fano}(1961)}]{Fano1961}%
  \BibitemOpen
  \bibfield  {author} {\bibinfo {author} {\bibfnamefont {U.}~\bibnamefont
  {Fano}},\ }\href@noop {} {\bibfield  {journal} {\bibinfo  {journal} {Am. J.
  Phys.}\ }\textbf {\bibinfo {volume} {29}},\ \bibinfo {pages} {539} (\bibinfo
  {year} {1961})}\BibitemShut {NoStop}%
\bibitem [{\citenamefont {Glauber}(1965)}]{Glauber1965}%
  \BibitemOpen
  \bibfield  {author} {\bibinfo {author} {\bibfnamefont {R.~J.}\ \bibnamefont
  {Glauber}},\ }in\ \href@noop {} {\emph {\bibinfo {booktitle} {Quantum
  {Optics} and {Electronics}}}},\ \bibinfo {editor} {edited by\ \bibinfo
  {editor} {\bibfnamefont {C.}~\bibnamefont {de~Witt}}, \bibinfo {editor}
  {\bibfnamefont {A.}~\bibnamefont {Blandin}}, \ and\ \bibinfo {editor}
  {\bibfnamefont {C.}~\bibnamefont {Cohen-Tannoudji}}}\ (\bibinfo  {publisher}
  {Gordon and Breach},\ \bibinfo {address} {New York},\ \bibinfo {year}
  {1965})\BibitemShut {NoStop}%
\bibitem [{\citenamefont {Hong}\ \emph {et~al.}(1987)\citenamefont {Hong},
  \citenamefont {Ou},\ and\ \citenamefont {Mandel}}]{Hong1987}%
  \BibitemOpen
  \bibfield  {author} {\bibinfo {author} {\bibfnamefont {C.~K.}\ \bibnamefont
  {Hong}}, \bibinfo {author} {\bibfnamefont {Z.~Y.}\ \bibnamefont {Ou}}, \ and\
  \bibinfo {author} {\bibfnamefont {L.}~\bibnamefont {Mandel}},\ }\href@noop {}
  {\bibfield  {journal} {\bibinfo  {journal} {Phys. Rev. Lett.}\ }\textbf
  {\bibinfo {volume} {59}},\ \bibinfo {pages} {2044} (\bibinfo {year}
  {1987})}\BibitemShut {NoStop}%
\bibitem [{Note1()}]{Note1}%
  \BibitemOpen
  \bibinfo {note} {By \enquote {mode}, we mean a single-particle wavefunction,
  whether in real space or some other space, which can be occupied by some
  number of identical particles.}\BibitemShut {Stop}%
\bibitem [{Note2()}]{Note2}%
  \BibitemOpen
  \bibinfo {note} {The requirement for four modes holds for systems of two
  particles. In the context of continuous variables, configurations involving
  only two modes can also lead to violations of Bell's inequalities (see, for
  instance, \cite {Wenger2003, Cavalcanti2011}).}\BibitemShut {Stop}%
\bibitem [{\citenamefont {Aspect}(1999)}]{Aspect:1999gm}%
  \BibitemOpen
  \bibfield  {author} {\bibinfo {author} {\bibfnamefont {A.}~\bibnamefont
  {Aspect}},\ }\href@noop {} {\bibfield  {journal} {\bibinfo  {journal}
  {Nature}\ }\textbf {\bibinfo {volume} {398}},\ \bibinfo {pages} {189}
  (\bibinfo {year} {1999})}\BibitemShut {NoStop}%
\bibitem [{\citenamefont {Aspect}(2015)}]{aspect2015viewpoint}%
  \BibitemOpen
  \bibfield  {author} {\bibinfo {author} {\bibfnamefont {A.}~\bibnamefont
  {Aspect}},\ }\href@noop {} {\bibfield  {journal} {\bibinfo  {journal}
  {Physics}\ }\textbf {\bibinfo {volume} {8}},\ \bibinfo {pages} {123}
  (\bibinfo {year} {2015})}\BibitemShut {NoStop}%
\bibitem [{Note3()}]{Note3}%
  \BibitemOpen
  \bibinfo {note} {A two-electron interference in four momentum modes was
  reported in Ref.~\cite {Waitz2016}, but without access to all four joint
  detection probabilities.}\BibitemShut {Stop}%
\bibitem [{\citenamefont {Penrose}(1996)}]{Penrose1996}%
  \BibitemOpen
  \bibfield  {author} {\bibinfo {author} {\bibfnamefont {R.}~\bibnamefont
  {Penrose}},\ }\href@noop {} {\bibfield  {journal} {\bibinfo  {journal} {Gen.
  Relat. Gravit.}\ }\textbf {\bibinfo {volume} {28}},\ \bibinfo {pages} {581}
  (\bibinfo {year} {1996})}\BibitemShut {NoStop}%
\bibitem [{\citenamefont {Horne}\ \emph {et~al.}(1989)\citenamefont {Horne},
  \citenamefont {Shimony},\ and\ \citenamefont {Zeilinger}}]{Horne:1989gi}%
  \BibitemOpen
  \bibfield  {author} {\bibinfo {author} {\bibfnamefont {M.~A.}\ \bibnamefont
  {Horne}}, \bibinfo {author} {\bibfnamefont {A.}~\bibnamefont {Shimony}}, \
  and\ \bibinfo {author} {\bibfnamefont {A.}~\bibnamefont {Zeilinger}},\
  }\href@noop {} {\bibfield  {journal} {\bibinfo  {journal} {Phys. Rev. Lett.}\
  }\textbf {\bibinfo {volume} {62}},\ \bibinfo {pages} {2209} (\bibinfo {year}
  {1989})}\BibitemShut {NoStop}%
\bibitem [{\citenamefont {Rarity}\ and\ \citenamefont
  {Tapster}(1990)}]{Rarity1990}%
  \BibitemOpen
  \bibfield  {author} {\bibinfo {author} {\bibfnamefont {J.~G.}\ \bibnamefont
  {Rarity}}\ and\ \bibinfo {author} {\bibfnamefont {P.~R.}\ \bibnamefont
  {Tapster}},\ }\href@noop {} {\bibfield  {journal} {\bibinfo  {journal} {Phys.
  Rev. Lett.}\ }\textbf {\bibinfo {volume} {64}},\ \bibinfo {pages} {2495}
  (\bibinfo {year} {1990})}\BibitemShut {NoStop}%
\bibitem [{\citenamefont {Kofler}\ \emph {et~al.}(2012)\citenamefont {Kofler},
  \citenamefont {Singh}, \citenamefont {Ebner}, \citenamefont {Keller},
  \citenamefont {Kotyrba},\ and\ \citenamefont {Zeilinger}}]{Kofler2012}%
  \BibitemOpen
  \bibfield  {author} {\bibinfo {author} {\bibfnamefont {J.}~\bibnamefont
  {Kofler}}, \bibinfo {author} {\bibfnamefont {M.}~\bibnamefont {Singh}},
  \bibinfo {author} {\bibfnamefont {M.}~\bibnamefont {Ebner}}, \bibinfo
  {author} {\bibfnamefont {M.}~\bibnamefont {Keller}}, \bibinfo {author}
  {\bibfnamefont {M.}~\bibnamefont {Kotyrba}}, \ and\ \bibinfo {author}
  {\bibfnamefont {A.}~\bibnamefont {Zeilinger}},\ }\href@noop {} {\bibfield
  {journal} {\bibinfo  {journal} {Phys. Rev. A}\ }\textbf {\bibinfo {volume}
  {86}},\ \bibinfo {pages} {032115} (\bibinfo {year} {2012})}\BibitemShut
  {NoStop}%
\bibitem [{\citenamefont {Lewis-Swan}\ and\ \citenamefont
  {Kheruntsyan}(2015)}]{Lewis-Swan2015}%
  \BibitemOpen
  \bibfield  {author} {\bibinfo {author} {\bibfnamefont {R.~J.}\ \bibnamefont
  {Lewis-Swan}}\ and\ \bibinfo {author} {\bibfnamefont {K.~V.}\ \bibnamefont
  {Kheruntsyan}},\ }\href@noop {} {\bibfield  {journal} {\bibinfo  {journal}
  {Phys. Rev. A}\ }\textbf {\bibinfo {volume} {91}},\ \bibinfo {pages} {052114}
  (\bibinfo {year} {2015})}\BibitemShut {NoStop}%
\bibitem [{Sup()}]{Supplemental}%
  \BibitemOpen
  \href@noop {} {}\bibinfo {note} {See Supplemental Material for a model of
  Bragg diffraction and details of other methods.}\BibitemShut {Stop}%
\bibitem [{\citenamefont {Clauser}\ \emph {et~al.}(1969)\citenamefont
  {Clauser}, \citenamefont {Horne}, \citenamefont {Shimony},\ and\
  \citenamefont {Holt}}]{Clauser1969}%
  \BibitemOpen
  \bibfield  {author} {\bibinfo {author} {\bibfnamefont {J.~F.}\ \bibnamefont
  {Clauser}}, \bibinfo {author} {\bibfnamefont {M.~A.}\ \bibnamefont {Horne}},
  \bibinfo {author} {\bibfnamefont {A.}~\bibnamefont {Shimony}}, \ and\
  \bibinfo {author} {\bibfnamefont {R.~A.}\ \bibnamefont {Holt}},\ }\href@noop
  {} {\bibfield  {journal} {\bibinfo  {journal} {Phys. Rev. Lett.}\ }\textbf
  {\bibinfo {volume} {23}},\ \bibinfo {pages} {880} (\bibinfo {year}
  {1969})}\BibitemShut {NoStop}%
\bibitem [{\citenamefont {Bonneau}\ \emph {et~al.}(2013)\citenamefont
  {Bonneau}, \citenamefont {Ruaudel}, \citenamefont {Lopes}, \citenamefont
  {Jaskula}, \citenamefont {Aspect}, \citenamefont {Boiron},\ and\
  \citenamefont {Westbrook}}]{Bonneau2013}%
  \BibitemOpen
  \bibfield  {author} {\bibinfo {author} {\bibfnamefont {M.}~\bibnamefont
  {Bonneau}}, \bibinfo {author} {\bibfnamefont {J.}~\bibnamefont {Ruaudel}},
  \bibinfo {author} {\bibfnamefont {R.}~\bibnamefont {Lopes}}, \bibinfo
  {author} {\bibfnamefont {J.-C.}\ \bibnamefont {Jaskula}}, \bibinfo {author}
  {\bibfnamefont {A.}~\bibnamefont {Aspect}}, \bibinfo {author} {\bibfnamefont
  {D.}~\bibnamefont {Boiron}}, \ and\ \bibinfo {author} {\bibfnamefont {C.~I.}\
  \bibnamefont {Westbrook}},\ }\href@noop {} {\bibfield  {journal} {\bibinfo
  {journal} {Phys. Rev. A}\ }\textbf {\bibinfo {volume} {87}},\ \bibinfo
  {pages} {061603} (\bibinfo {year} {2013})}\BibitemShut {NoStop}%
\bibitem [{\citenamefont {Schellekens}\ \emph {et~al.}(2005)\citenamefont
  {Schellekens}, \citenamefont {Hoppeler}, \citenamefont {Perrin},
  \citenamefont {Gomes}, \citenamefont {Boiron}, \citenamefont {Aspect},\ and\
  \citenamefont {Westbrook}}]{Schellekens2005}%
  \BibitemOpen
  \bibfield  {author} {\bibinfo {author} {\bibfnamefont {M.}~\bibnamefont
  {Schellekens}}, \bibinfo {author} {\bibfnamefont {R.}~\bibnamefont
  {Hoppeler}}, \bibinfo {author} {\bibfnamefont {A.}~\bibnamefont {Perrin}},
  \bibinfo {author} {\bibfnamefont {J.~V.}\ \bibnamefont {Gomes}}, \bibinfo
  {author} {\bibfnamefont {D.}~\bibnamefont {Boiron}}, \bibinfo {author}
  {\bibfnamefont {A.}~\bibnamefont {Aspect}}, \ and\ \bibinfo {author}
  {\bibfnamefont {C.~I.}\ \bibnamefont {Westbrook}},\ }\href {\doibase
  10.1126/science.1118024} {\bibfield  {journal} {\bibinfo  {journal}
  {Science}\ }\textbf {\bibinfo {volume} {310}},\ \bibinfo {pages} {648}
  (\bibinfo {year} {2005})}\BibitemShut {NoStop}%
\bibitem [{\citenamefont {Lopes}\ \emph {et~al.}(2015)\citenamefont {Lopes},
  \citenamefont {Imanaliev}, \citenamefont {Aspect}, \citenamefont {Cheneau},
  \citenamefont {Boiron},\ and\ \citenamefont {Westbrook}}]{Lopes2015}%
  \BibitemOpen
  \bibfield  {author} {\bibinfo {author} {\bibfnamefont {R.}~\bibnamefont
  {Lopes}}, \bibinfo {author} {\bibfnamefont {A.}~\bibnamefont {Imanaliev}},
  \bibinfo {author} {\bibfnamefont {A.}~\bibnamefont {Aspect}}, \bibinfo
  {author} {\bibfnamefont {M.}~\bibnamefont {Cheneau}}, \bibinfo {author}
  {\bibfnamefont {D.}~\bibnamefont {Boiron}}, \ and\ \bibinfo {author}
  {\bibfnamefont {C.~I.}\ \bibnamefont {Westbrook}},\ }\href {\doibase
  10.1038/nature14331} {\bibfield  {journal} {\bibinfo  {journal} {Nature}\
  }\textbf {\bibinfo {volume} {520}},\ \bibinfo {pages} {66} (\bibinfo {year}
  {2015})}\BibitemShut {NoStop}%
\bibitem [{\citenamefont {Wenger}\ \emph {et~al.}(2003)\citenamefont {Wenger},
  \citenamefont {Hafezi}, \citenamefont {Grosshans}, \citenamefont
  {Tualle-Brouri},\ and\ \citenamefont {Grangier}}]{Wenger2003}%
  \BibitemOpen
  \bibfield  {author} {\bibinfo {author} {\bibfnamefont {J.}~\bibnamefont
  {Wenger}}, \bibinfo {author} {\bibfnamefont {M.}~\bibnamefont {Hafezi}},
  \bibinfo {author} {\bibfnamefont {F.}~\bibnamefont {Grosshans}}, \bibinfo
  {author} {\bibfnamefont {R.}~\bibnamefont {Tualle-Brouri}}, \ and\ \bibinfo
  {author} {\bibfnamefont {P.}~\bibnamefont {Grangier}},\ }\href@noop {}
  {\bibfield  {journal} {\bibinfo  {journal} {Phys. Rev. A}\ }\textbf {\bibinfo
  {volume} {67}},\ \bibinfo {pages} {012105} (\bibinfo {year}
  {2003})}\BibitemShut {NoStop}%
\bibitem [{\citenamefont {Cavalcanti}\ \emph {et~al.}(2011)\citenamefont
  {Cavalcanti}, \citenamefont {Brunner}, \citenamefont {Skrzypczyk},
  \citenamefont {Salles},\ and\ \citenamefont {Scarani}}]{Cavalcanti2011}%
  \BibitemOpen
  \bibfield  {author} {\bibinfo {author} {\bibfnamefont {D.}~\bibnamefont
  {Cavalcanti}}, \bibinfo {author} {\bibfnamefont {N.}~\bibnamefont {Brunner}},
  \bibinfo {author} {\bibfnamefont {P.}~\bibnamefont {Skrzypczyk}}, \bibinfo
  {author} {\bibfnamefont {A.}~\bibnamefont {Salles}}, \ and\ \bibinfo {author}
  {\bibfnamefont {V.}~\bibnamefont {Scarani}},\ }\href@noop {} {\bibfield
  {journal} {\bibinfo  {journal} {Phys. Rev. A}\ }\textbf {\bibinfo {volume}
  {84}},\ \bibinfo {pages} {022105} (\bibinfo {year} {2011})}\BibitemShut
  {NoStop}%
\bibitem [{\citenamefont {Waitz}\ \emph {et~al.}(2016)\citenamefont {Waitz},
  \citenamefont {Metz}, \citenamefont {Lower}, \citenamefont {Schober},
  \citenamefont {Keiling}, \citenamefont {Pitzer}, \citenamefont {Mertens},
  \citenamefont {Martins}, \citenamefont {Viefhaus}, \citenamefont {Klumpp},
  \citenamefont {Weber}, \citenamefont {Schmidt-B{\"o}cking}, \citenamefont
  {Schmidt}, \citenamefont {Morales}, \citenamefont {Miyabe}, \citenamefont
  {Rescigno}, \citenamefont {McCurdy}, \citenamefont {Mart{\'\i}n},
  \citenamefont {Williams}, \citenamefont {Sch{\"o}ffler}, \citenamefont
  {Jahnke},\ and\ \citenamefont {D{\"o}rner}}]{Waitz2016}%
  \BibitemOpen
  \bibfield  {author} {\bibinfo {author} {\bibfnamefont {M.}~\bibnamefont
  {Waitz}}, \bibinfo {author} {\bibfnamefont {D.}~\bibnamefont {Metz}},
  \bibinfo {author} {\bibfnamefont {J.}~\bibnamefont {Lower}}, \bibinfo
  {author} {\bibfnamefont {C.}~\bibnamefont {Schober}}, \bibinfo {author}
  {\bibfnamefont {M.}~\bibnamefont {Keiling}}, \bibinfo {author} {\bibfnamefont
  {M.}~\bibnamefont {Pitzer}}, \bibinfo {author} {\bibfnamefont
  {K.}~\bibnamefont {Mertens}}, \bibinfo {author} {\bibfnamefont
  {M.}~\bibnamefont {Martins}}, \bibinfo {author} {\bibfnamefont
  {J.}~\bibnamefont {Viefhaus}}, \bibinfo {author} {\bibfnamefont
  {S.}~\bibnamefont {Klumpp}}, \bibinfo {author} {\bibfnamefont
  {T.}~\bibnamefont {Weber}}, \bibinfo {author} {\bibfnamefont
  {H.}~\bibnamefont {Schmidt-B{\"o}cking}}, \bibinfo {author} {\bibfnamefont
  {L.}~\bibnamefont {Schmidt}}, \bibinfo {author} {\bibfnamefont
  {F.}~\bibnamefont {Morales}}, \bibinfo {author} {\bibfnamefont
  {S.}~\bibnamefont {Miyabe}}, \bibinfo {author} {\bibfnamefont
  {T.}~\bibnamefont {Rescigno}}, \bibinfo {author} {\bibfnamefont
  {C.}~\bibnamefont {McCurdy}}, \bibinfo {author} {\bibfnamefont
  {F.}~\bibnamefont {Mart{\'\i}n}}, \bibinfo {author} {\bibfnamefont
  {J.}~\bibnamefont {Williams}}, \bibinfo {author} {\bibfnamefont
  {M.}~\bibnamefont {Sch{\"o}ffler}}, \bibinfo {author} {\bibfnamefont
  {T.}~\bibnamefont {Jahnke}}, \ and\ \bibinfo {author} {\bibfnamefont
  {R.}~\bibnamefont {D{\"o}rner}},\ }\href@noop {} {\bibfield  {journal}
  {\bibinfo  {journal} {Phys. Rev. Lett.}\ }\textbf {\bibinfo {volume} {117}},\
  \bibinfo {pages} {083002} (\bibinfo {year} {2016})}\BibitemShut {NoStop}%
\end{thebibliography}%

\end{document}